*Type of the Paper (Article)*

# An adaptive neuro-fuzzy model for attitude estimation and control a 3 DOF system

Xin Wang[1,*], Seyed Mehdi Abtahi [2], Mahmood Chahari[3] and Tianyu Zhao[4,*]

1. Department of Kinesiology, Shenyang Sport University, Shenyang 110102, China；wangxin@syty.edu.cn
2. Department of Mechanical engineering, University of Illinois at Chicago, Chicago, USA; sabtah2@uic.edu
3. Department of Mechanical engineering, State University of New York at Binghamton, 4400 Vestal Parkway, Binghamton, NY 13902, USA; mchahar1@binghamton.edu
4. Key Laboratory of Structural Dynamics of Liaoning Province, College of Sciences, Northeastern University, Shenyang 110819, China; zhaotianyu@mail.neu.edu.cn
* Correspondence: zhaotianyu@mail.neu.edu.cn ; wangxin@syty.edu.cn

**Abstract:** In recent decades, one of the scientists' main concerns has been to improve the accuracy of satellite attitude, regardless of the expense. The obvious result is that a large number of control strategies have been used to address this problem. In this study, an adaptive neuro-fuzzy integrated (ANFIS) satellite attitude estimation and control system was developed. The controller is trained with the data provided by an optimal controller. A pulse modulator is used to generate the right ON/OFF commands of the thruster actuator. To evaluate the performance of the AN-FIS controller in closed-loop simulation, an ANFIS observer is used to estimate the attitude and angular velocities of the satellite using magnetometer, sun sensor and data gyro data. In addition, a new ANFIS system will be proposed and evaluated that can jointly control and estimate the system. The performance of the ANFIS controller is compared to the optimal PID controller in a Monte Carlo simulation with different initial conditions, disturbance and noise. The results show that the ANFIS controller can surpass the optimal PID controller in several aspects, including time and smoothness. In addition, the ANFIS estimator is examined and the results demonstrate the high ability of this designated observers. Both the control and estimation phases are simulated by a single ANFIS subsystem, taking into account the high capacity of ANFIS, and the results of using the ANFIS model are demonstrated.

## 1. Introduction

Satellite attitude control plays a significant role in most space missions. Therefore, the development of an accurate and stable controller is an essential part of conducting a space mission (Inamori et al., 2011; Ismail and Varatharajoo, 2010). The most advanced satellite attitude control techniques use the concept of quaternion feedback (Fossen, 2002; Tavakoli and Assadian, 2018). Different linear and nonlinear attitude control strategies based on quaternion feedback were considered (Abtahi and Sharifi, 2020; Wu et al., 2017). The quaternion feedback approach also used to stabilize the attitude of microsatellites (Kristiansen and Nicklasson, 2005).

In recent years, a vast majority of control techniques have been used to closely control satellite attitude in the presence of uncertainty and disturbance (Aleksandrov et al., 2018; Kang et al., 2017). Li et al. (Li et al., 2017) proposed a robust finite time control al-gorithm for controlling satellite attitude in uncertainty. Xiao et al. (Xiao et al., 2016) developed a control with a simple structure to perform an attitude maneuver in case of disturbances and uncertain inertia parameters. Vatankhahghadim and Damaren (Vatankhahghadim and Damaren, 2017) have adopted the passivity rate for the hybrid attitude control of a spacecraft using magnetic torques and thrusters.

Several different types of optimal controllers have been used to enhance the satellite attitude control system. Zhang Fan et al. (Fan et al., 2002) In order to improve the accuracy of a small satellite, an attempt was made to optimize the attitude control model. In another study, the optimal magnetic attitude control of is studied (Wisniewski and Markley, 1999). Arantes et al. (Arantes et al., 2009) has tried to analyze and design a reaction thruster attitude controller and then improve the performance of the control subsystem. All these optimal controls inevitably led to a specific mathematical model, leading to inappropriate behavior compared to external pulses in the comparison simulation state. More importantly, an optimal controller may not be able to perform the task in the presence of uncertainties.

The adaptive control method is one of the most powerful models that can deal with the problem of uncertainty. In this regard, Wen et al. (Wen et al., 2017) used adaptive attitude controller to control agile spacecraft. An adaptive controller was used to control the satellite attitude by solar radiation from Lee and Singh (Lee and Singh, 2014). In another research, they (Lee and Singh, 2009) a non-insecure, equivalent, adaptive satellite attitude controller that uses the sun's radiation pressure. All of these adaptive control logics are model-based, and although they are able to work accurately with uncertainties, they cannot work with different dynamic models. Determining the satellite attitude has been the main concern of many studies in recent decades (Cao and Li, 2016; Zeng et al., 2014; Zhang et al., 2013). Kouyama et al. (Kouyama et al., 2017) used an image fitting method to determine the satellite attitude, which of course follows an exact map projection. They used this method together with the classic onboard sensors. Wu et al. (Wu et al., 2017) proposed a method by which the problem of orientation based on a single sensor observation can be solved.

The enormous ability of fuzzy logic to solve various mathematical problems of modeling, control and estimation is undeniable (Daley and Gill, 1986), which used self-organizer fuzzy logic controller (SOC) to control a flexible satellite that has significant dynamic coupling of the axes that cannot be modelled easily. Mukherjee et al. (Mukherjee et al., 2017) used fuzzy logic to control the attitude of earth-pointing satellites, in which they used the genetic algorithm to optimize the performance of their proposed nonlinear fuzzy PID controller. In the other research, Huo et al. (Huo et al., 2016) proposed an adaptive fuzzy fault tolerance attitude control for a rigid spacecraft. In recent years, fuzzy logic has been used for a variety of satellite attitude estimation purposes (Zhong et al., 2015). For example, adaptive fuzzy fault tolerance control for rigid spacecraft attitude maneuvers was studied by Ran et al. (Ran et al., 2016). Sun et al. (Sun et al., 2017) used an adaptive fuzzy estimator for spacecraft attitude determination.

In this paper, an ANFIS (adapted neuro-fuzzy inference system) (Jang, 1993) controller is introduced to control and estimate the satellite attitude. Given the tremendous ability of ANFIS to control and estimate, there has been no research on the integrated control and estimation of estimation and control of satellite attitude using ANFIS. The proposed feature of the proposed model is the elimination of interphase (sensors equations and equations used to calculate quaternion errors). This, in turn, eliminates systematic errors and noise that are unavoidable in classical approaches. Consequently, the ANFIS control method is overly applicable in terms of measurement noise, model uncertainty, and external disturbance.

The organization of this paper is as follows. First, a summary of the satellite attitude dynamics is given. A brief overview of the optimal PID controller design for control systems will then be given. Next, the general ANFIS structure and the learning algorithms will be discussed. Subsequently, structures of ANFIS controller and satellite attitude estimator are given. Finally, an ANFIS integrated control and estimation subsystem is introduced to reduce the complexity of the control system. The usefulness of this model is examined by comparing the proposed model results with those of the classical controller.

## 2. Modelling of System

### 2.1. Satellite Dynamics Model

In this section, we introduce equations of motion of a satellite with Euler equation and quaternion kinematics. The Euler equation of the rigid body satellite attitude around its principal axes coordinates is (Wie, 1998):

$$\begin{aligned} I_1\dot{\omega}_1 &= M_{c1} + M_{d1} - (I_3 - I_2)\omega_2\omega_3 \\ I_2\dot{\omega}_2 &= M_{c2} + M_{d2} - (I_1 - I_3)\omega_1\omega_3 \\ I_3\dot{\omega}_3 &= M_{c3} + M_{d3} - (I_2 - I_1)\omega_2\omega_1 \end{aligned} \quad (1)$$

where $\omega_1, \omega_2$ and $\omega_3$ are the elements of angular velocity vector of satellite. $I_1$, $I_2$, and $I_3$ are the moments of inertia about the principal axes. $M_c$ and $M_d$ are control and disturbance moments, respectively, which are expressed in the body frame.

For kinematic representation, the quaternion vector $\bar{q} = (q_1, q_2, q_3, q_4)^T$ is utilized, which is defined as follows:

$$\begin{bmatrix} q_1 \\ q_2 \\ q_3 \end{bmatrix} = \sin\frac{\theta}{2} \begin{bmatrix} e_1 \\ e_2 \\ e_3 \end{bmatrix}, \qquad q_4 = \sin\frac{\theta}{2} \quad (2)$$

where $\theta$ is the rotation angle about the Euler axis $\bar{e} = (e_1, e_2, e_3)$. The kinematic differential equations for quaternions are as follows:

$$\begin{bmatrix} \dot{q}_1 \\ \dot{q}_2 \\ \dot{q}_3 \\ \dot{q}_4 \end{bmatrix} = \frac{1}{2}\begin{bmatrix} 0 & \omega_3 & -\omega_2 & \omega_1 \\ -\omega_3 & 0 & \omega_1 & \omega_2 \\ \omega_2 & -\omega_1 & 0 & \omega_3 \\ -\omega_1 & -\omega_2 & -\omega_3 & 0 \end{bmatrix}\begin{bmatrix} q_1 \\ q_2 \\ q_3 \\ q_4 \end{bmatrix} \quad (3)$$

### 2.2. Measurements

The sun sensor and the magnetometer are the sensors used in this study to estimate the setting. In order to simulate the magnetometer sensor (magnetic field), height, latitude, longitude date, are considered as inputs and the magnetic field vector is calculated as inertia frame $\bar{B}^I$ using IGRF11 model (Finlay et al., 2010). Then, the magnetic field is transformed into the body frame $\bar{B}^B$ including a random white noise $\bar{n}_B$:

$$\bar{B}^B = C_I^B \bar{B}^I + \bar{n}_B \quad (4)$$

The rotation matrix $C_I^B$, is calculated using the quaternion vector as follows:

$$C_I^B = \begin{bmatrix} 1-2(q_2^2+q_3^2) & 2(q_1q_2+q_3q_4) & 2(q_1q_3+q_2q_4) \\ 2(q_2q_1+q_3q_4) & 1-2(q_1^2+q_3^2) & 2(q_2q_3+q_1q_4) \\ 2(q_3q_1+q_2q_4) & 2(q_3q_2+q_1q_4) & 1-2(q_1^2+q_2^2) \end{bmatrix} \quad (5)$$

The attitude measurement needs only the direction of the magnetic field which can be calculated as:

$$\bar{u}_B^B = \bar{B}^B/|\bar{B}^B|, \quad (6)$$

The sun vector direction in inertial frame $\bar{u}_S^I$ can be found by the following formulation (Vallado, 2001):

$$\begin{aligned}JD &= 367\text{year} - \text{INT}\left[\frac{7\left(\text{year} + \text{INT}\left(\frac{\text{month}+9}{12}\right)\right)}{4}\right] \\ &+ \text{INT}\left(\frac{275\text{month}}{9}\right) + \text{day} + 1721013.5 \\ &+ \frac{\left(\frac{\text{second}}{60} + \text{minute}\right)}{60} + \text{hour}}{24} \\ T &= (JD - 2451545.0)/36525 \\ \lambda_M &= 280.4606184° + 36000.77005361T \\ M &= 357.5277233° + 35999.05034T \\ \lambda_{ecliptic} &= \lambda_M + 1.914666471° \sin(M) \\ &\quad + 0.019994643\sin(2M) \\ \varepsilon &= 23.439291° - 0.0130042T \\ \bar{u}_S^I &= \left(\cos(\lambda_{ecliptic})\cos(\varepsilon)\sin(\lambda_{ecliptic})\sin(\varepsilon)\sin(\lambda_{ecliptic})\right)^T\end{aligned} \quad (7)$$

where JD is Julian Day based on the date and time (year, month, day, hour, minute and second), $T$ is the Julian centuries, $\lambda_M$ is mean longitude of the sun, $M$ is the mean anomaly of the sun, $\lambda_{ecliptic}$ is the ecliptic longitude of the sun, and $\varepsilon$ is the tilt angle of the Earth rotation axis.

Similar to the magnetometer, the output of the sun sensor as the direction of the sun vector in body frame $\bar{u}_S^B$ can be estimated as follows:

$$\bar{u}_S^B = C_I^B \bar{u}_S^I + \bar{n}_S, \quad (8)$$

Furthermore, to provide the angular velocity measurements, a three-axis rate-gyro with random white noise is used.

## 3. Adaptive Neural Fuzzy Inference System

### 3.1. Fuzzy Logic

Most traditional tools for modeling, thinking, and arithmetic are crisp, deterministic, and precise in character, so yes or no type instead of more or less type. In conventional dual logic, for example, a statement may be true or false and nothing in between. For the first time, L.A.Zadeh (Zadeh, 1965) proposed a fuzzy logic that contained "true", "false," and "partially true." He emphasized that real situations are often not clear and deterministic and cannot be described accurately.

A fuzzy control system is based on fuzzy logic that analyzes input values in the form of logical variables that assume continuous values between 0 and 1. The fuzzy logic was first used by Mamdani and Assilian (Mamdani and Assilian, 1975) in the engineering problem. Rather than designing algorithms that explicitly define the control action as a function of the control input variables, the developer of a fuzzy controller writes rules that associate the input variables with the control variables through expressions of linguistic variables. After all rules have been defined, the control process begins with the calculation of all rule sequences. Then the consequences are summarized into a fuzzy set that describes the possible control actions.

*3.2. ANFIS*

In general, the fuzzy control logic has two main approaches; 1) Mamdani (Mamdani and Assilian, 1975) and 2) Takagi-Sugeno (Takagi and Sugeno, 1985). The basis of ANFIS as an adaptive network-based fuzzy system is the Takagi-Sugeno fuzzy system method (Jang, 1993). Its inference system corresponds to a set of fuzzy IF-THEN rules that have a learning ability to approximate non-linear functions. ANFIS is a combination of neural networks and fuzzy systems. However, ANFIS has become a very powerful simulation method that uses both fuzzy and neural network methods. Recently, ANFIS modeling has become widespread in various space missions (Gupta et al., 2017; O. S. Hanafy, 2014; Ting and Bo, 2013).

The most important feature of the ANFIS controller is the ability to handle a free model system that allows the use of real data and, more importantly, the design of a controller based on the provided real data. The other considerable superiority of the ANFIS system is the required number of input variables for control and estimation. Simplicity of modeling compared to classical modeling, along with the superiority of this method in the presence of noise and uncertainty compared to PID controllers, which makes our proposed model more acceptable.

ANFIS has 5 layers (As shown in **Figure 1**) as follows:

Layer 1: Define membership function of input variables.

$$O_{1,i} = \mu_{A_i}(x) \quad for \ i = 1,2$$
$$O_{1,i} = \mu_{B_{i-2}}(x) \quad for \ i = 3,4$$

Layer 2: Product of the membership function for each input.

$$O_{2,i} = \omega_i = \mu_{A_i}(x)\, \mu_{B_i}(x) \quad i = 1,2$$

Layer 3: Normalize the output of layer 2.

$$O_{3,i} = \bar{\omega}_i = \frac{\omega_i}{\omega_1 + \omega_2} \quad i = 1,2$$

Layer 4: The output of this layer is:

$$O_{4,i} = \bar{\omega}_i f_i = \bar{\omega}_i (p_i x + q_i y + r_i)$$

Layer 5: The output of this layer is the summation of all outputs in layer 4.

$$O_{5,i} = \sum \bar{\omega}_i f_i = \frac{\sum \omega_i f_i}{\sum \omega_i}$$

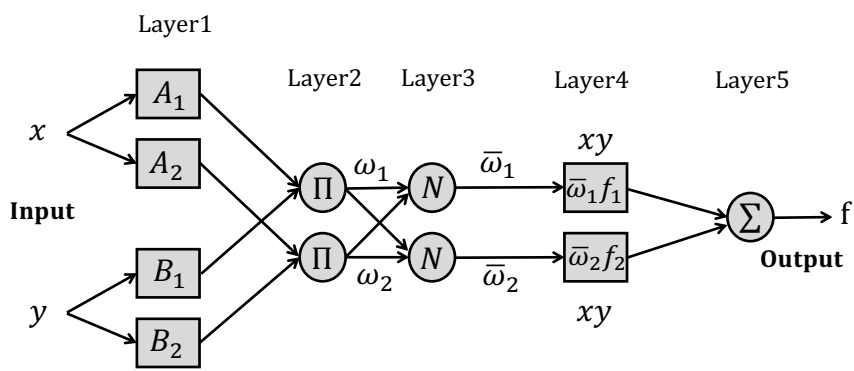

**Figure 1.** ANFIS Structure.

### 3.3. Hybrid Learning Algorithm

Least Square Gradient Reduction is used to train the ANFIS system (locating the membership function parameters) the pattern between the inputs and the output data provided by an optimal PID controller.

Each learning level is divided into two parts. In the forward stage, the inputs and outputs of each layer are calculated and optimal coefficients are provided. Then in the reverse stage the parameters of the ANFIS system are updated.

### 3.4. Optimal PID Controller

The control moment vector by using PID controller can be calculated as:

$$M_c = K_p q_e + K_d \omega + K_q \int q_e dt + K_\omega \int \omega dt, \tag{9}$$

where $q_e$ is the quaternion error and can be obtained from the following equation (Wie, 1998):

$$\begin{bmatrix} \bar{q}_e \\ q_4 \end{bmatrix} = \begin{bmatrix} q_{1e} \\ q_{2e} \\ q_{3e} \\ q_{4e} \end{bmatrix} = \begin{bmatrix} q_{4c} & q_{3c} & -q_{2c} & -q_{1c} \\ -q_{3c} & q_{4c} & q_{1c} & -q_{2c} \\ q_{2c} & -q_{1c} & q_{4c} & -q_{3c} \\ q_{1c} & q_{2c} & q_{3c} & q_{4c} \end{bmatrix} \begin{bmatrix} q_1 \\ q_2 \\ q_3 \\ q_4 \end{bmatrix}, \tag{10}$$

where $q_c$s are the quaternions of the command attitude.

The control gains $(K_p, K_d, K_q)$ in Eq. (7) are optimized in order to minimize the following cost function:

$$J = \int \left( \sum_{i=1}^{3} |\omega_i| + \sum_{i=1}^{3} |q_{e_i}| \right) dt, \tag{11}$$

By considering the following constraint as:

$$|M_c| \le M_{c_{max}}, \tag{12}$$

Consequently, this constraint guarantees the appropriate signal command to input the modulator for ON-OFF command of the thrusters with torque $M_{c_{max}}$.

## 4. ANFIS Controller and Estimator

### 4.1. ANFIS Controller

The aim of this training is to model an optimal PID controller as close as possible. The control input variables are angular velocity and quaternion errors, and the control output variable is the control torque $M_C$ (as shown in **Figure 2**).

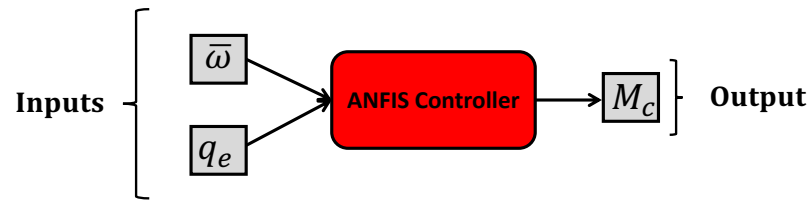

**Figure 2.** Block diagram of ANFIS controller.

After the input and output variables are supplied by a system with PID controller, the collection of this data is repeated several times, taking into account 15 different initial conditions (each simulation for 20 seconds with 0.01 second sampling time). This means that the initial quaternions and initial angular velocities are changed to provide a wide range of data for ANFIS learning. Thereafter, the ANFIS controller training process begins and the ANFIS system learns the path from the inputs to the outputs. Now the ANFIS controller can work with all initial conditions.

*4.2. ANFIS Estimator*

In this study, we have utilized sun sensor and magnetometer outputs to estimate attitude. Thus, data for ANFIS estimation learning from these two sensors is provided both in the body (sensor) and in the inertia frame (calculation) (as shown in **Figure 3**). Several different scenarios are considered to provide a large database for learning the ANFIS estimator.

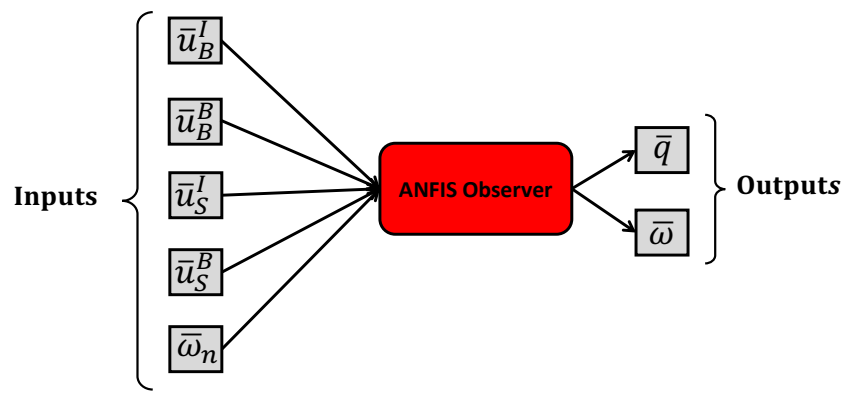

**Figure 3.** Block diagram of ANFIS observer.

*4.3. Combined Control and Estimation using ANFIS*

In this study, both the ANFIS estimator and the ANFIS controller are used in a closed loop. The nesting simulations show the performance of these two ANFIS subsystems working simultaneously (as shown in **Figure 4**).

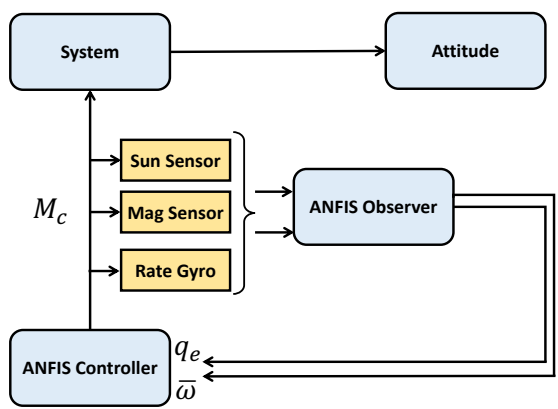

**Figure 4.** Block diagram of the combined ANFIS observer and controller.

## 4.4. Integrated Control and Estimation using ANFIS

As mentioned in the introduction, the main purpose of this study is to evaluate the performance of an ANFIS system as a combination of estimator and controller instead of two separate subsystems (ANFIS estimator and ANFIS controller). As shown in **Figure 5**, for the proposed ANFIS subsystem, input variables are the inputs of the estimator (sensor data) and output variables are the outputs of the controller (control torque). In fact, the ANFIS integrated control and estimation subsystem receives data read by the sun sensor and the magnetic sensor as input variables and then passes the control torque directly to the system dynamics as shown in **Figure 6**).

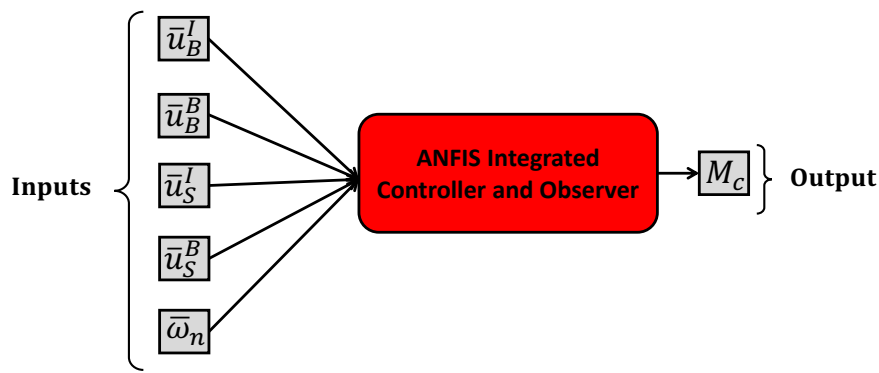

**Figure 5.** Block diagram of the integrated ANFIS controller and observer.

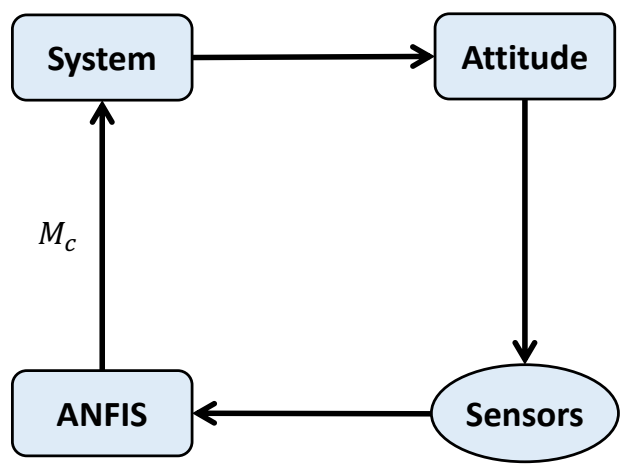

**Figure 6.** Block diagram of control system using ANFIS integrated controller and observer.

## 5. Evaluation of ANFIS control and estimation

To study the performance of attitude estimation and control of satellite using ANFIS, a satellite with the moments of inertia that presented in Error! Reference source not found. is considered. For all simulations, the final simulation duration time is selected to be 20 seconds, and the sampling time for estimation is 0.01 sec. The system initial conditions and the desire attitude are provided as in Error! Reference source not found..

**Table 1.** Nominal and indeterminate moments of inertia (in Kg.m2).

|  | $I_x$ | $I_y$ | $I_z$ |
|---|---|---|---|
| Moment of Inertia | 1.5 | 2.6 | 3 |
| Moments of Inertia in case of uncertainty | 2.5 | 4 | 3.3 |

**Table 2.** Sample initial condition (this initial condition is not in the training set).

|  | $\omega_x$(Rad/s) | $\omega_y$(Rad/s) | $\omega_z$(Rad/s) | $\phi$(deg) | $\theta$(deg) | $\varphi$(deg) |
|---|---|---|---|---|---|---|
| Initial condition | 0.0125 | 0.05 | 0.075 | 10 | 5 | 10 |
| Desired condition | 0 | 0 | 0 | 5 | 0 | 0 |

### 5.1. ANFIS Performance Comparison

As the simulations are for stabilization of satellite attitude on zero condition, the most important characteristics of the results are the settling time of control, the control effort (fuel consumption) and the steady state error. Therefore, these characteristics are considered as the criteria for comparison of the results.

The comparison of time histories of control moments for PID and ANFIS in presence of noise and uncertainty are shown in **Figure 7** and Figure 9, respectively. The trajectory of the Euler angles are also presented in **Figure 8** and **Figure 10**. From **Figure 7**, it can be seen that the PID controller is noisy and the ANFIS controller produces smoother control actions. Moreover, the trajectory of the attitude angles using PID controller has larger over-shoot values.

The numerical results of comparison of the ANFIS controller and PID controller are provided in **Error! Reference source not found.** with/without noise and uncertainty. It is clear from this table that the fuel consumption of ANFIS controller is 5% lower than PID even if there is no uncertainty and/or measurement noise. The presence of noise and uncertainty induces more fuel consumption (14% and 9%, respectively).

The settling time with 1% error is listed in **Error! Reference source not found.** for both controllers. The improvement of settling time using ANFIS over the PID is more obvious. In some cases, the settling time of ANFIS is almost half the PID, which is very important in space systems.

**Table 3.** Fuel consumption of PID and ANFIS controllers (in N.M.S).

|  | X axe | Y axe | Z axe | Total |
|---|---|---|---|---|
| **Without noise and uncertainty** | | | | |
| **ANFIS** | 0.1311 | 0.3956 | 0.7208 | 1.2475 |
| **PID** | 0.1287 | 0.4282 | 0.7485 | 1.3054 |
| **Considering noise** | | | | |
| **ANFIS** | 0.1732 | 0.4117 | 0.6925 | 1.2774 |
| **PID** | 0.1983 | 0.4719 | 0.8126 | 1.4830 |
| **Considering uncertainty** | | | | |
| **ANFIS** | 0.1910 | 0.6030 | 0.7891 | 1.5831 |
| **PID** | 0.1992 | 0.7048 | 0.8343 | 1.7383 |

**Table 4.** Settling time for 1% error for satellite Euler angles using PID and ANFIS controllers (in second).

| | X axe | Y axe | Z axe |
|---|---|---|---|
| **Without noise and uncertainty** | | | |
| **ANFIS** | 7.23 | 4.87 | 8.88 |
| **PID** | 9.62 | 8.82 | 9.51 |
| **Considering noise** | | | |
| **ANFIS** | 6.4 | 6.62 | 4.94 |
| **PID** | 9.34 | 8.68 | 9.46 |
| **Considering uncertainty** | | | |
| **ANFIS** | 4.59 | 11.38 | 10.9 |
| **PID** | 9.84 | 10.65 | 10 |

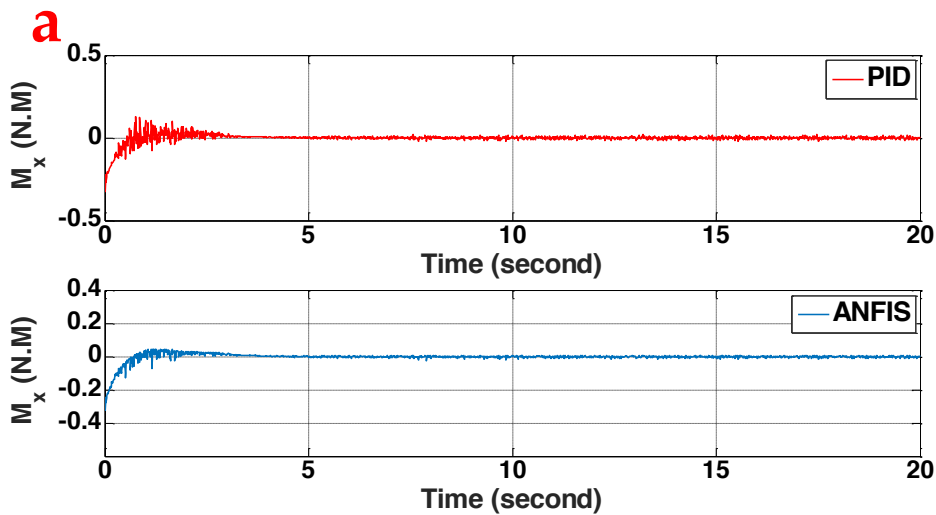

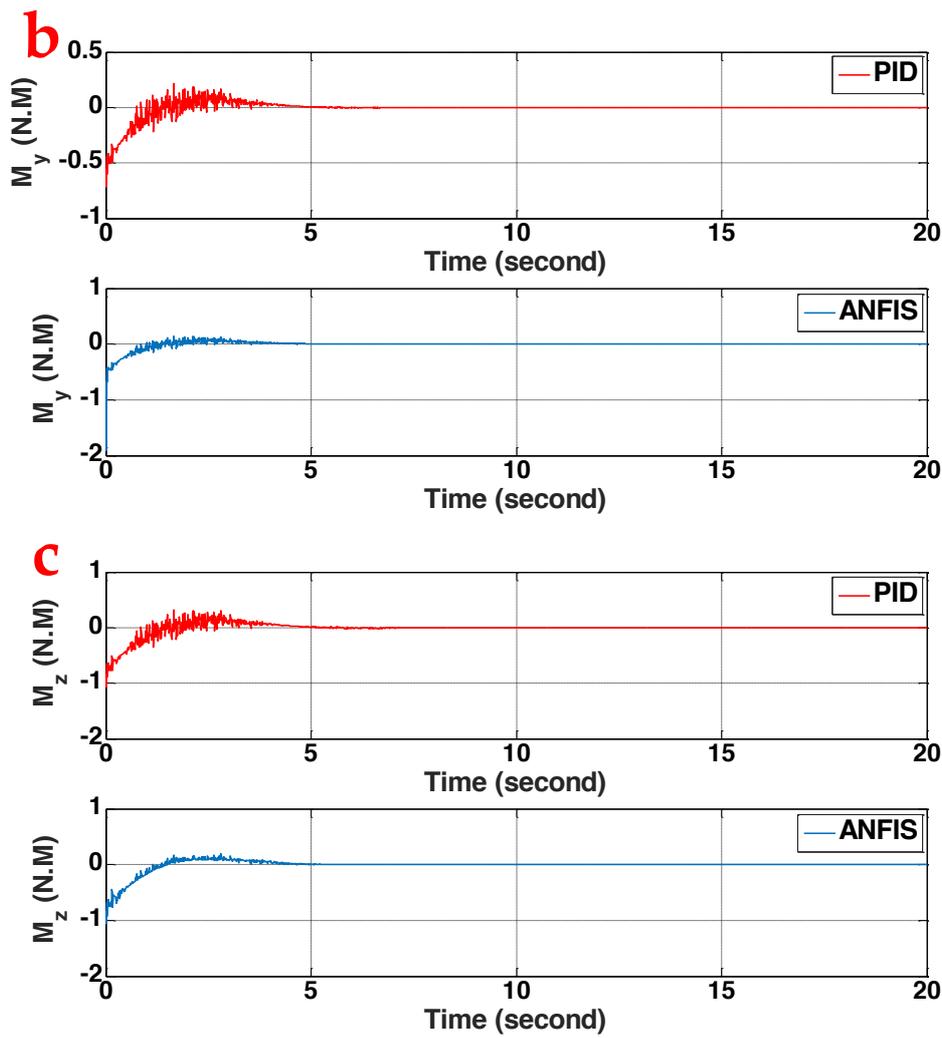

**Figure 7.** Comparison of control moment in (a) X (b) Y and (c) Z directions using ANFIS and PID in presence of noise.

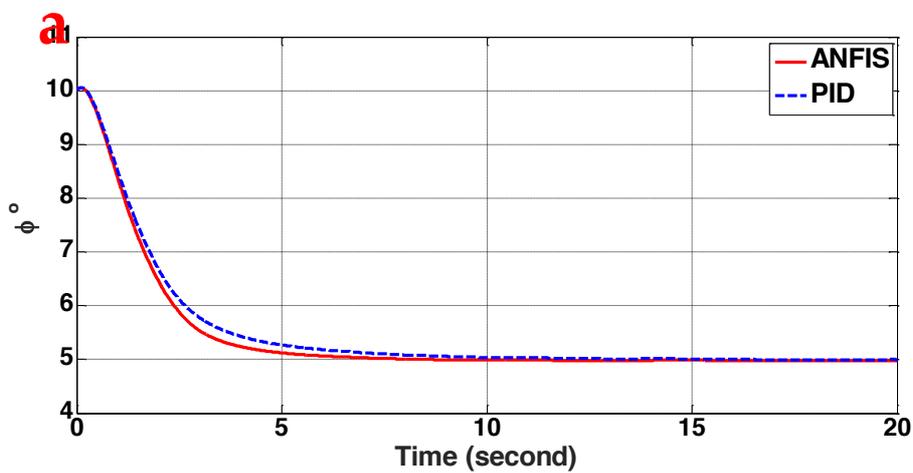

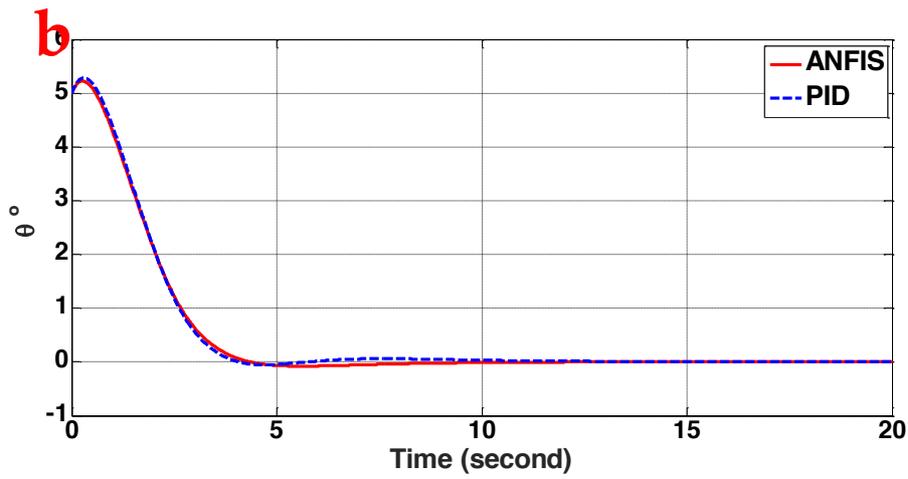

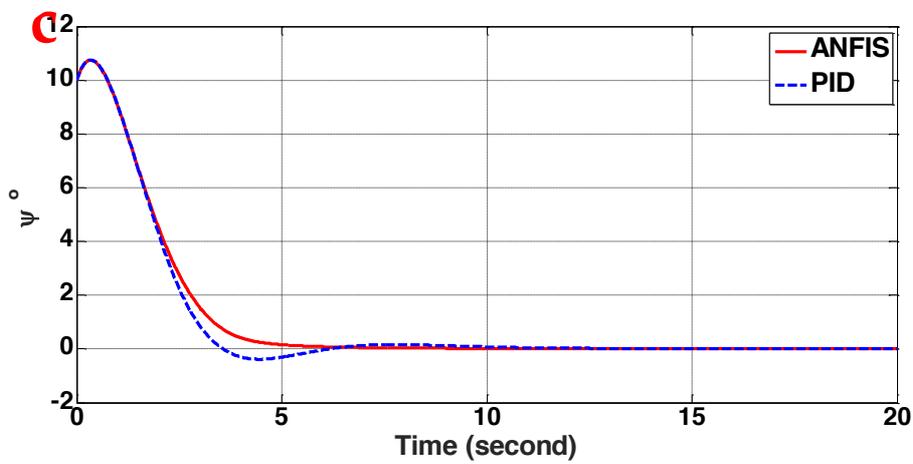

**Figure 8.** Comparison of Euler angles (a) φ (b) θ and (c) ψ using ANFIS and PID controller in presence of noise.

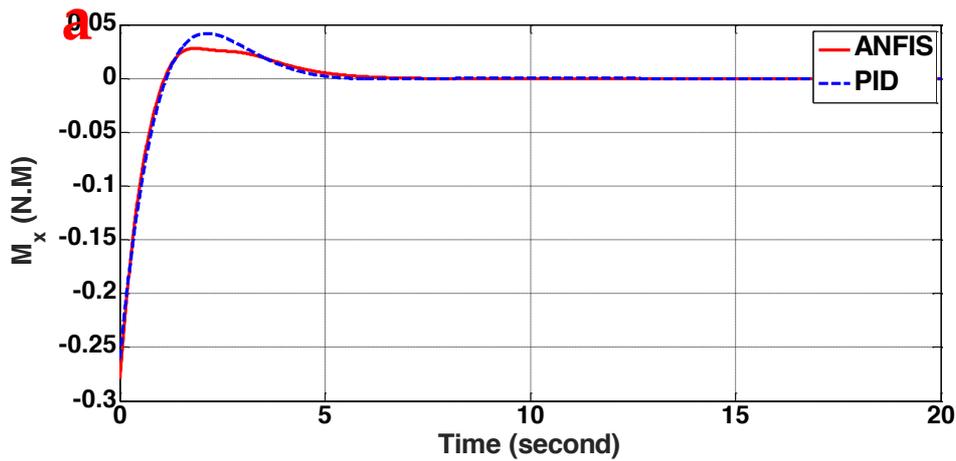

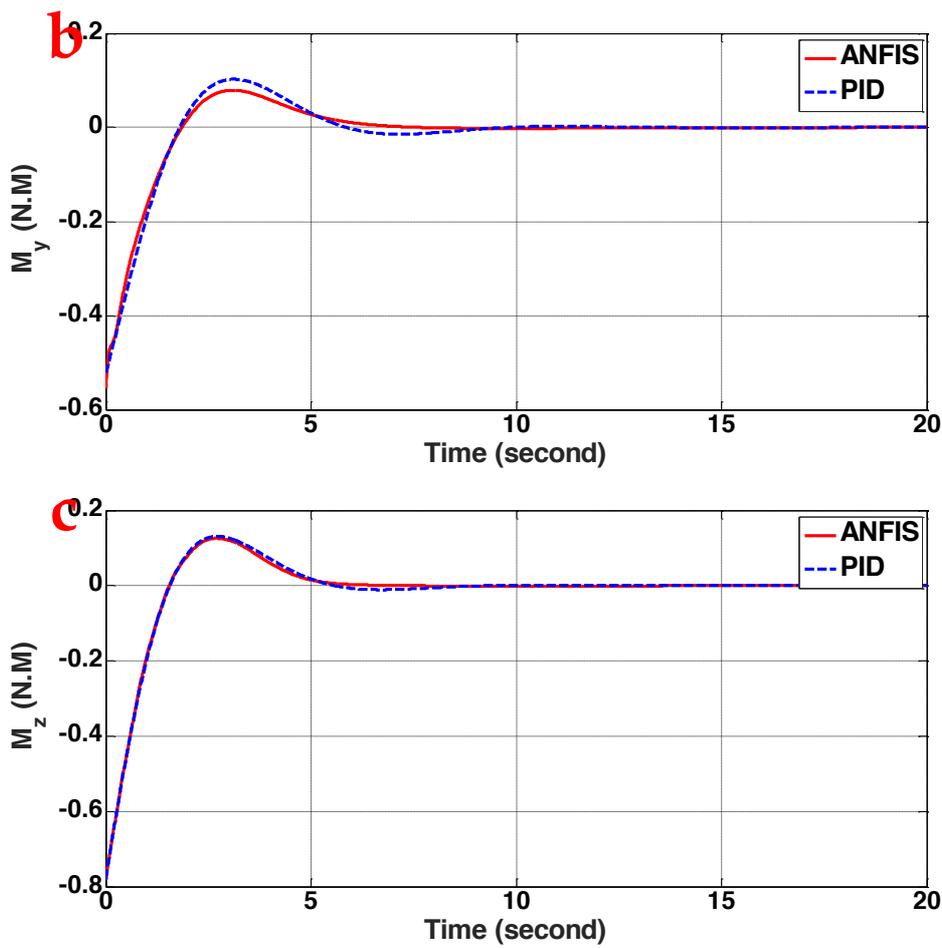

**Figure 9.** Comparison of control moment in (a) X (b) Y and (c) Z directions using ANFIS and PID in presence of uncertainty.

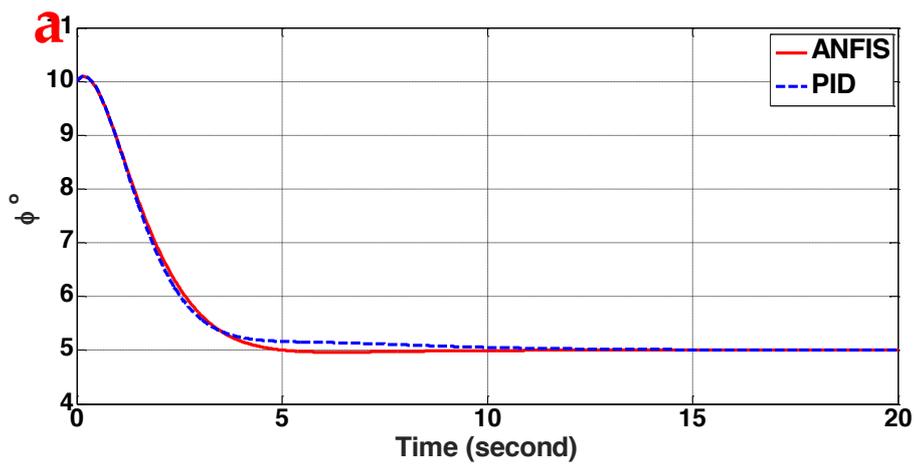

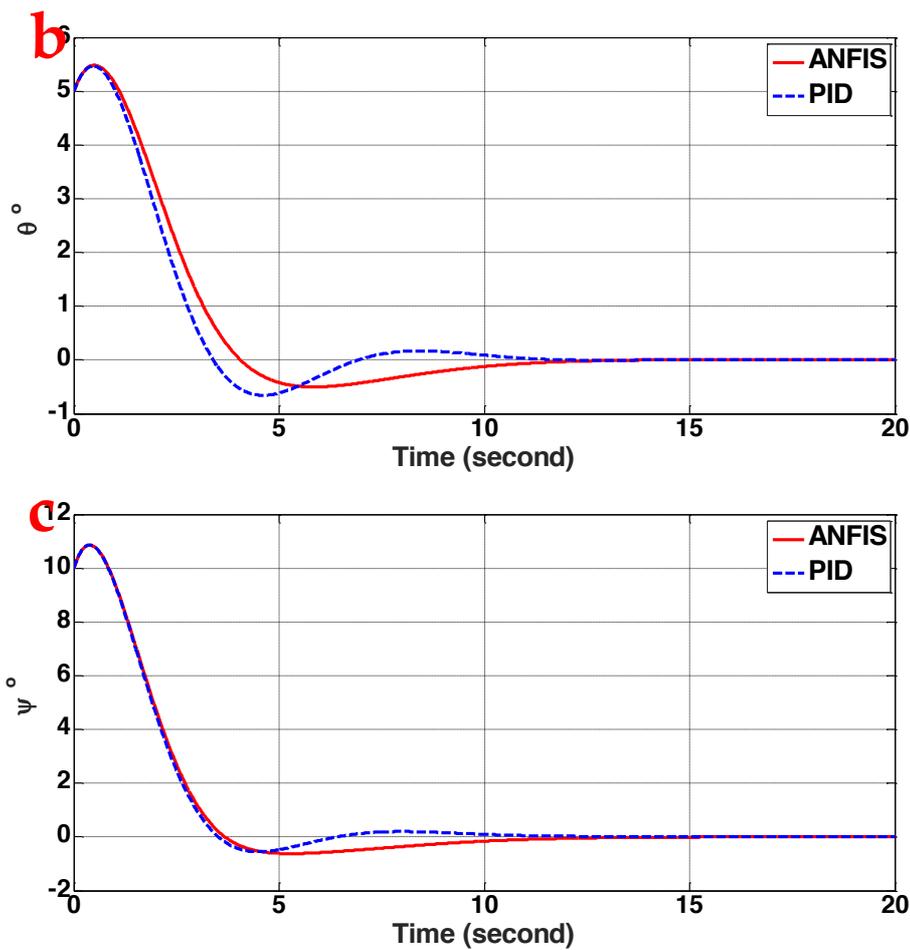

**Figure 10.** Comparison of Euler angles (a) φ (b) θ and (c) ψ using ANFIS and PID controller in presence of uncertainty.

### 5.2. Command Modulation

To evaluate the ANFIS controller results for the real thruster actuator, the control moments should be converted to ON-OFF commands. Thus, a PWPF (Pulse-Width Pulse-Frequency) modulator is used to transform the continuous control moment command to the ON-OFF commands. The trajectory of the attitude angles and the thruster commands are shown in **Figure 11** and **Figure 12**, respectively. It is clear that the limit of thrust results in slower approach to the final attitudes. However, the results are acceptable considering both uncertainty and measurement noise exist.

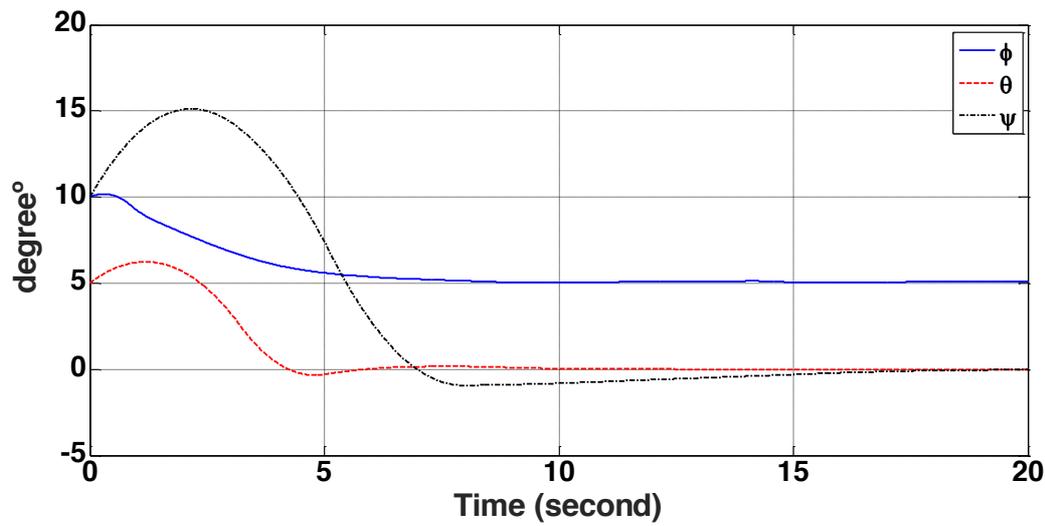

**Figure 11.** Euler angles time history using PWPF modulator

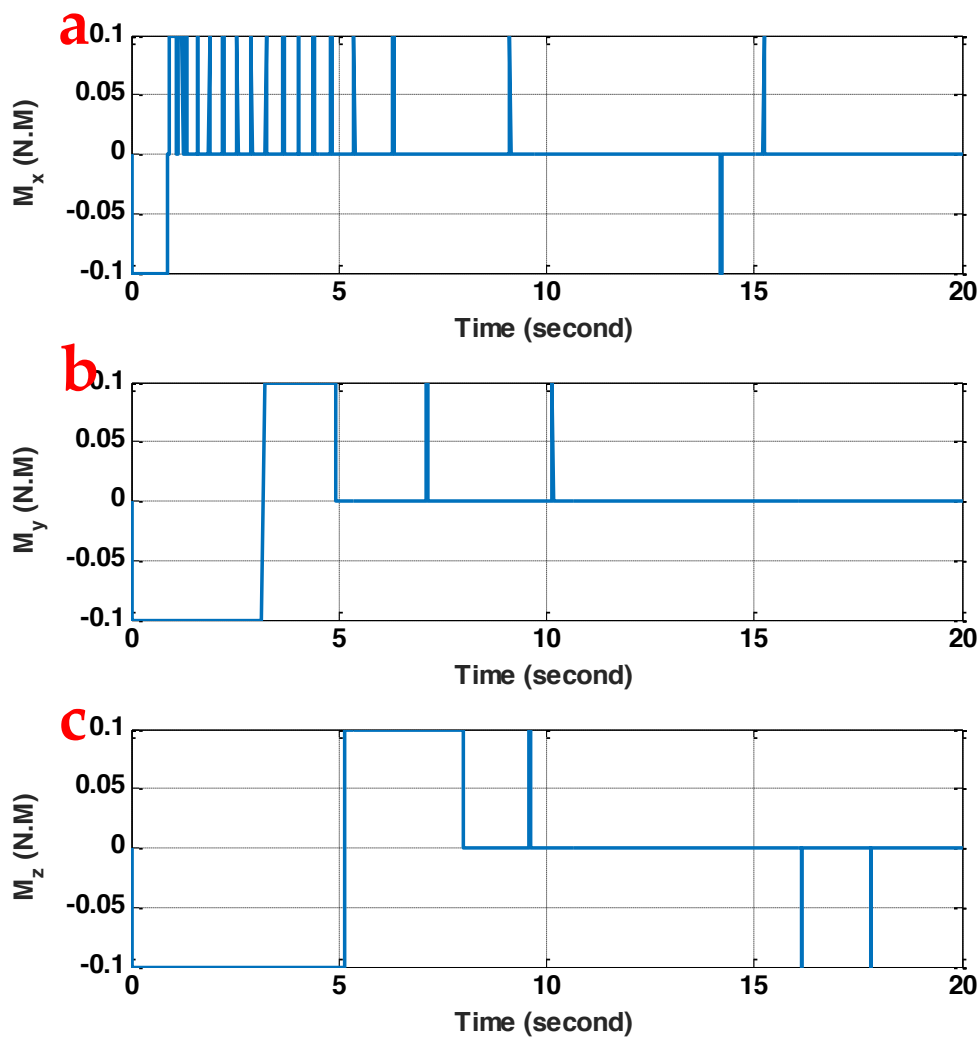

**Figure 12.** The control moment of Thruster in (a) X, (b) Y, and (c) Z axis.

*5.3. Monte Carlo Simulation*

To analyze the robustness of the proposed integrated ANFIS estimator and controller, a Monte-Carlo simulation is done. A random initial condition (between -15 and +15 degrees) are considered in addition to the random noise and random uncertainty (I 1 Kg.m2). The attitude control error of Euler angles of time=20sec are shown for each Monte-Carlo simulation and the average and 3σ (standard deviation) until each iteration is shown in **Figure 13** for 200 iteration. The maximum control error is less than 0.02 degrees, which is considerably low.

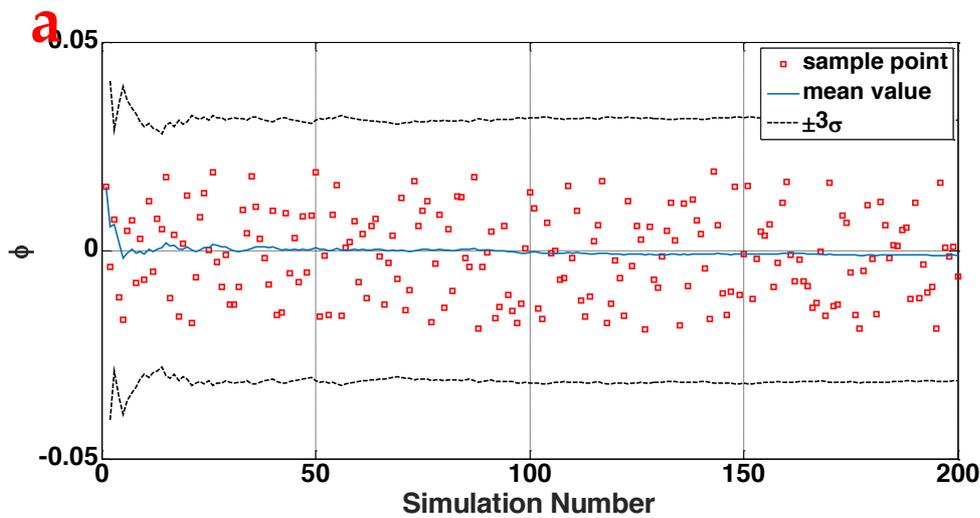

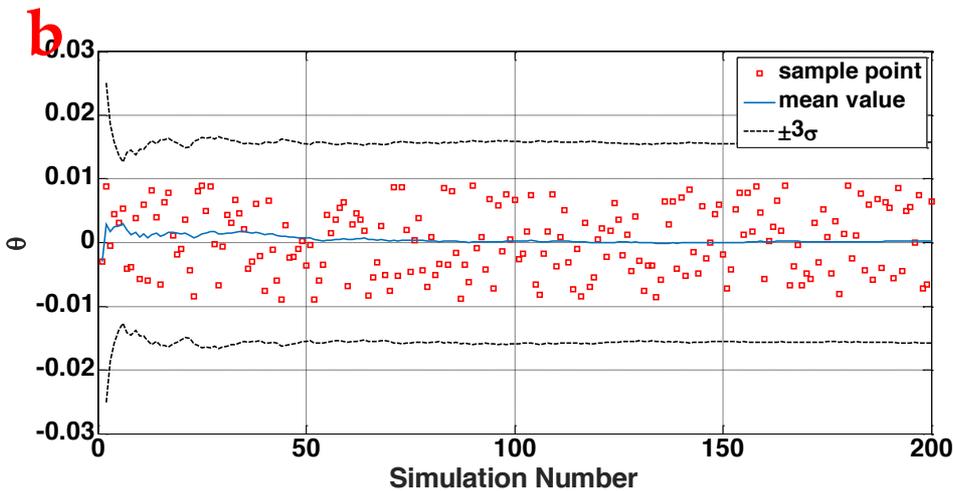

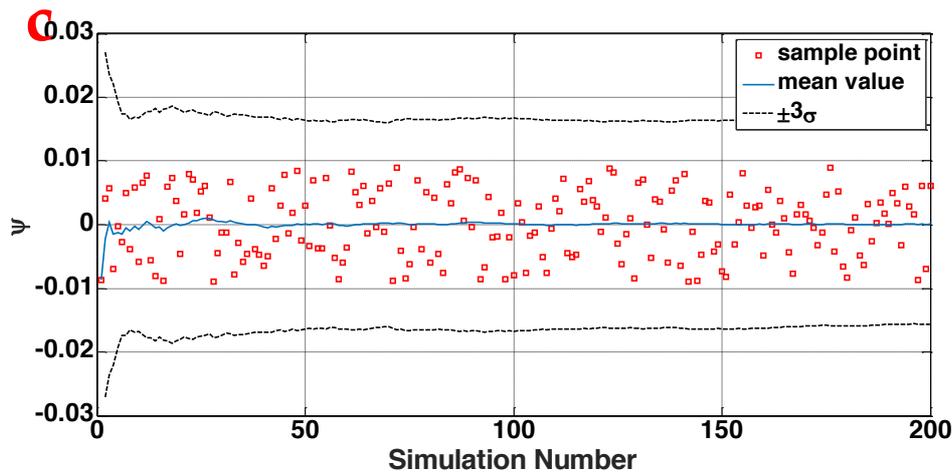

(c)

**Figure 13.** Monte-Carlo simulation error for (a) φ, (b) θ, and (c) ψ.

## 6. Conclusions

In this study, an ANFIS (adaptive neuro-fuzzy inference system) controller and estimator was proposed in order to estimate and control the attitude of a satellite. ANFIS controller was trained using an optimal PID. The other significant ability of the proposed system was estimating necessary states accurately via ANFIS observer. To train the observer, a satellite in several different conditions (noise and uncertainty) were considered. The performance of the ANFIS controller and estimator was compared with the PID controller in the presence of uncertainties and noises.

Comparing the performance of PID and ANFIS controller shows that the ANFIS controller consumed less control effort (fuel) in all situations (noise and uncertainty). In addition, ANFIS controller outputs behave smoother and reach stability in a shorter time interval than PID controller. Likewise, system outputs (control angles) are smoother and reach to desire angles faster. Using ANFIS estimator in system showed that despite of its simple design, it could estimate the states even in presence of uncertainty. Results of using synchronous control and estimation ANFIS simulator show that although both stages (control and estimation) are done in one-step, the performance of the integrated system is similar to the combined controller and estimator. Due to the proved abilities of the ANFIS controller and observer, it can be concluded that it can be able to work with black box systems. It means that the determined dynamic is not essential, which makes it possible to be used for unknown space bodies (like space debris) as well as fast parameter varying space objects (like space robots and manipulators)

**Funding:** The Key Special Project of the National Key Research and Development Program "Technical Winter Olympics" (2018YFF0300502 and 2021YFF0306400).

**Conflicts of Interest:** The authors declare no conflict of interest.